\documentclass[11pt]{article}
\usepackage{epic}
\usepackage{amsfonts}
          \def\cH{{\cal H}}

\def\ZZ{{\mathbb Z}}
\def\one#1{#1^{\raise5pt\hbox{$\scriptstyle\!\!\!\!1$}}\,{}}
\def\two#1{#1^{\raise5pt\hbox{$\scriptstyle\!\!\!\!2$}}\,{}}
\def\sn{{\rm{sn}}}
\def\cn{{\rm{cn}}}
\def\dn{{\rm{dn}}}
\begin{document}
%%%%%%%%%%%%%%%%%%%%%%%%%%%%%%%%%%%%%%%%%%%%%%%%%%%%%%%%%%%%%%%%%%%%%%%%%%%%%%
\begin{center} 
{\LARGE \textsf{
Integrable XYZ Model with Staggered 
\\[4mm]
Anisotropy Parameter}} 

\vspace{36pt}
{\large
{\bf D.~Arnaudon\footnote{e-mail:{\sl arnaudon@lapp.in2p3.fr}}},
{\bf D.~Karakhanyan}\footnote{e-mail:{\sl karakhan@lx2.yerphi.am}},
{\bf M.~Mirumyan}\footnote{e-mail:{\sl mkhitar@moon.yerphi.am}},\\

{\bf A.~Sedrakyan}\footnote{e-mail:{\sl sedrak@lx2.yerphi.am}}
{\bf P.~Sorba\footnote{e-mail:{\sl sorba@lapp.in2p3.fr}}}\\ 
}

\vspace{30pt}

\emph{Laboratoire d'Annecy-le-Vieux de Physique Th\'eorique LAPTH}
\\
\emph{CNRS, UMR 5108, associ\'ee {\`a} l'Universit\'e de Savoie}
\\
\emph{BP 110, F-74941 Annecy-le-Vieux Cedex, France}
\\
\emph{Yerevan Physics Institute, Br.~Alikhanian str.2, Yerevan, Armenia}
\vspace{36pt}

\vfill
{\bf Abstract}
\end{center}

We apply to the $XYZ$ model the technique of construction of integrable 
models with staggered parameters, presented recently for the $XXZ$ case
in \cite{APSS}. The solution of modified
Yang--Baxter equations is found and the corresponding integrable  
zig-zag ladder Hamiltonian is calculated. The result is coinciding
with the $XXZ$ case in the appropriate limit.

\vfill
\rightline{LAPTH-874/01}
\rightline{hep-th/0111233}
\rightline{November 2001}

\newpage
\pagestyle{plain}

\section{Introduction}
\setcounter{equation}{0}
\indent

Recently a technique was proposed \cite{APSS} which allows 
to construct a zig-zag ladder type of integrable models on the basis
of known integrable chain models. It was successfully
realised on models with basic $sl(n)$ symmetries; $sl(2)$ case
(XXZ model) in \cite{APSS}, $sl(3)$ case (anisotropic $t-J$ model)
in \cite{TS} and general $sl(n)$ case in the  \cite{ASSS}.
The main element of this construction is the possibility to stagger
the parameters of known integrable $R$-matrix along chain and time
directions. Together with alternation of the anisotropy parameter
$\Delta$ \cite{APSS, TS} the staggered shift of the spectral
parameter $u$ by some additional model parameter $\theta$ was
also introduced, which caused the appearance of the next to nearest
neighbour interaction terms in the resulting local Hamiltonian.

The motivation for considering this type of integrable models
is based on the observation that the phenomenological
Chalker-Coddington model \cite{CC} for edge excitations in the Hall
effect after its reformulation as a lattice field theory \cite{AS1}
and calculation of disorder over random phases becomes a Hubbard type
model with staggered disposition of $R$-matrices \cite{AS2}. Therefore
it is meaningful  to start the investigation of staggered models
from simple cases.

In this letter we generalise the construction of the staggered $XXZ$ 
model and construct a staggered integrable $XYZ$ model. 
The corresponding generalised Yang--Baxter Equations ($YBE$),
which are the condition of integrability \cite{B,FT}, have a
solution. Therefore, one can define a model on a two rung ladder, the
Hamiltonian of which is presented here.
%%%%%%%%%%%%%%%%%%%%%%%%%%%%%%%%%%%%%%%%%%%%%%%%%%%%%%%%%%%%%%%%%%%
%%%%%%%%%%%%%%%%%%%%%%%%%%%%%%%%%%%%%%%%%%%%%%%%%%%%%%%%%%%%%%%%%%%%
\section{Basic Definitions and Staggered $YBE$'s}
\setcounter{equation}{0}

\indent

We start by writing down the modification of basic
ingredients
of the Algebraic Bethe Ansatz ($ABA$) technique \cite{B, FT}
appropriate for our purposes.
 
Let us consider $\ZZ_2$ graded quantum $V_{j,\rho}(v)$ 
(with $j=1,.....N$ as a chain index) and 
auxiliary $V_{a,\sigma}(u)$ spaces, where $\rho, \sigma =0,1$ are
the grading indices. Consider
$R$-matrices, which act on the direct product
of  spaces $V_{a,\sigma}(u)$ and $ V_{j,\rho}(v)$, $(\sigma,\rho =0,1)$,
mapping them on the intertwined direct product of 
$V_{a,\bar{\sigma}}(u)$ and $ V_{j,\bar{\rho}(v)}$ with the complementary
$\bar{\sigma}=(1-\sigma)$, $\bar{\rho}=(1-\rho)$ indices
\begin{equation}
\label{R1}
R_{aj,\sigma \rho}\left( u,v\right):\quad V_{a,\sigma}(u)\otimes 
V_{j,\rho}(v)\rightarrow V_{j,\bar{\rho}}(v)\otimes V_{a,\bar{\sigma}}(u).  
\end{equation}

It is convenient to introduce 
two transmutation operations $\iota_1$
and $\iota_2$ with the property $\iota_1^2=\iota_2^2=id$ 
for the quantum and auxiliary spaces
correspondingly, and to define the operators $R_{aj,\sigma\rho}$ as 
follows
\begin{eqnarray}
\label{R2}
R_{aj,00}&\equiv& R_{aj},\qquad R_{aj,01}\equiv R_{aj}^{\iota_1},\nonumber\\
R_{aj,10}&\equiv& R_{aj}^{\iota_2},\qquad R_{aj,11}\equiv R_{aj}^{\iota_1 
\iota_2}.
\end{eqnarray}

The introduction of the $\ZZ_2$ grading of quantum spaces 
in time direction means
that we have now two monodromy operators $T_{\rho}, \rho=0,1$,
which act on the space $V_{\rho}(u)=\prod_{j=1}^N V_{j,\rho}(u)$
by mapping it on $V_{\bar{\rho}}(u)=\prod_{j=1}^N V_{j,\bar{\rho}}(u)$
\begin{equation}
\label{T}
T_\rho(v,u) \qquad : V_\rho(u) \rightarrow V_{\bar{\rho}}(u), \qquad \qquad 
\rho=0,1.
\end{equation}

It is clear now, that the monodromy operator of the model, which is defined
by translational invariance by two steps in the time direction and
which determines the partition function, is the product of two
monodromy operators 
\begin{equation}
\label{TT}
T(v,u) = T_0(v,u) T_1(v,u).
\end{equation}

The $\ZZ_2$ grading of auxiliary spaces along the chain direction means
that the $T_0(u,v)$ and $T_1(u,v)$ monodromy matrices are defined
according to the following staggered product
of the $R_{aj}(v,u)$ and $\bar{R}_{aj}^{\iota_2}(v,u)$ matrices:
\begin{eqnarray}
\label{T1}
T_1(v,u)=\prod_{j=1}^N R_{a,2j-1}(v,u)
\bar{R}_{a,2j}^{\iota_2}(v,u)\nonumber\\
T_0(v,u)=\prod_{j=1}^N \bar{R}_{a,2j-1}^{\iota_1}(v,u)
R_{a,2j}^{\iota_1 \iota_2}(v,u),
\end{eqnarray}
where the notation $\bar{R}$ denotes a 
different parametrisation of the $R(v,u)$-matrix via spectral
parameters
$v$ and $u$ and can be considered as an operation
over $R$ with property $\bar{\bar{R}}= R$.
For the integrable models where the intertwiner matrix $R(v-u)$
simply depends
on the difference of the spectral parameters $v$ and $u$ 
this operation is a shift of its argument $u$:
\begin{eqnarray}
 \label{RR}
 \bar{R}(u)=R(\bar u), \qquad \bar{u}=\theta-u,
 \end{eqnarray}
where $\theta$ is an additional model parameter. We will consider
this case in this paper.

As it is well known in Bethe Ansatz technique \cite{B, FT}, the sufficient
condition for the commutativity of transfer matrices $\tau(u)=
Tr T(u)$ with different spectral parameters is the YBE. For our
case we have a set of two equations \cite{APSS}
\begin{eqnarray}
  \label{eq:YBE1}
  R_{12}(u,v) \bar{R}_{13}^{\iota_1}(u) R_{23}(v)=
  R_{23}^{\iota_1}(v) \bar{R}_{13}(u) \tilde{R}_{12}(u,v)
\end{eqnarray}
and
\begin{eqnarray}
  \label{eq:YBE2}
  \tilde{R}_{12}(u,v) R_{13}^{\iota_1 \iota_2}(u) 
  \bar{R}_{23}^{\iota_2}(v)=
  \bar{R}_{23}^{\iota_1 \iota_2}(v) R_{13}^{\iota_2}(u) R_{12}(u,v) \;.
\end{eqnarray}

%%%%%%%%%%%%%%%%%%%%%%%%%%%%%%%%%%%%%%%%%%%%%%%%%%%%%%%%%%%%%%%%%%
\section{Staggered $XYZ$ Heisenberg chain}
\setcounter{equation}{0}

\indent
The $R$-matrix of the ordinary $XYZ$ Heisenberg model can be 
represented as
\begin{equation}
\label{R1/2}
R(u)=\left(
\begin{array}{cccc}
a(u)&&&d(u)\\
&b(u)&c(u)&\\
&c(u)&b(u)&\\
d(u)&&&a(u)
\end{array}\right)
\end{equation}

Inputting this expression of $R(u)$ into the $YBE$ (\ref{eq:YBE1})
we obtain the following three sets of equations on
$a(u), b(u), c(u), d(u)$. The first set is:

$$
a(u,v)a^{\iota_1}(\bar u)a(v)+d(u,v)c^{\iota_1}(\bar u)d(v)=
a^{\iota_1}(v)a(\bar u)\tilde a(u,v)+d^{\iota_1}(v)c(\bar u)\tilde d(u,v),
$$
$$
a(u,v)b^{\iota_1}(\bar u)b(v)+d(u,v)d^{\iota_1}(\bar u)c(v)=
b^{\iota_1}(v)b(\bar u)\tilde a(u,v)+c^{\iota_1}(v)d(\bar u)\tilde d(u,v),
$$
\begin{equation}
\label{i}
b(u,v)b^{\iota_1}(\bar u)a(v)+c(u,v)d^{\iota_1}(\bar u)d(v)=
a^{\iota_1}(v)b(\bar u)\tilde b(u,v)+d^{\iota_1}(v)d(\bar u)\tilde c(u,v),
\end{equation}
$$
b(u,v)a^{\iota_1}(\bar u)b(v)+c(u,v)c^{\iota_1}(\bar u)c(v)=
b^{\iota_1}(v)a(\bar u)\tilde b(u,v)+c^{\iota_1}(v)c(\bar u)\tilde c(u,v),
$$
which are satisfied automatically in case of trivial $\iota_1$
operation, as it happened in ordinary $XYZ$ model. 

The second set
$$
a(u,v)b^{\iota_1}(\bar u)c(v)+d(u,v)d^{\iota_1}(\bar u)b(v)=
c^{\iota_1}(v)a(\bar u)\tilde b(u,v)+b^{\iota_1}(v)c(\bar u)\tilde c(u,v),   
$$
$$ 
b(u,v)a^{\iota_1}(\bar u)c(v)+c(u,v)c^{\iota_1}(\bar u)b(v)=
c^{\iota_1}(v)b(\bar u)\tilde a(u,v)+b^{\iota_1}(v)d(\bar u)\tilde d(u,v),
$$
$$ 
a(u,v)c^{\iota_1}(\bar u)a(v)+d(u,v)a^{\iota_1}(\bar u)d(v)=
c^{\iota_1}(v)a(\bar u)\tilde c(u,v)+b^{\iota_1}(v)c(\bar u)\tilde b(u,v),
$$
\begin{equation}
\label{i2} 
c(u,v)a^{\iota_1}(\bar u)c(v)+b(u,v)c^{\iota_1}(\bar u)b(v)=
a^{\iota_1}(v)c(\bar u)\tilde a(u,v)+d^{\iota_1}(v)a(\bar u)\tilde d(u,v),
\end{equation}
$$ 
b(u,v)c^{\iota_1}(\bar u)c(v)+c(u,v)a^{\iota_1}(\bar u)b(v)=
a^{\iota_1}(v)b(\bar u)\tilde c(u,v)+d^{\iota_1}(v)d(\bar u)\tilde b(u,v),
$$
$$ 
b(u,v)d^{\iota_1}(\bar u)d(v)+c(u,v)b^{\iota_1}(\bar u)a(v)=
b^{\iota_1}(v)a(\bar u)\tilde c(u,v)+c^{\iota_1}(v)c(\bar u)\tilde b(u,v),
$$
is reducing to staggered $XXZ$ models equations \cite{APSS}
in case of $d(u)=0$. 
Finally we have a third set of equations
$$ 
a(u,v)a^{\iota_1}(\bar u)d(v)+d(u,v)c^{\iota_1}(\bar u)a(v)=
a^{\iota_1}(v)d(\bar u)\tilde c(u,v)+d^{\iota_1}(v)b(\bar u)\tilde b(u,v),
$$
$$ 
b(u,v)b^{\iota_1}(\bar u)d(v)+c(u,v)d^{\iota_1}(\bar u)a(v)=
a^{\iota_1}(v)c(\bar u)\tilde d(u,v)+d^{\iota_1}(v)a(\bar u)\tilde a(u,v),
$$
$$ 
b(u,v)d^{\iota_1}(\bar u)a(v)+c(u,v)b^{\iota_1}(\bar u)d(v)=
b^{\iota_1}(v)d(\bar u)\tilde a(u,v)+c^{\iota_1}(v)b(\bar u)\tilde d(u,v),
$$
\begin{equation}
\label{i3}
a(u,v)d^{\iota_1}(\bar u)b(v)+d(u,v)b^{\iota_1}(\bar u)c(v)=
a^{\iota_1}(v)d(\bar u)\tilde b(u,v)+d^{\iota_1}(v)b(\bar u)\tilde c(u,v),
\end{equation}
$$ 
a(u,v)c^{\iota_1}(\bar u)d(v)+d(u,v)a^{\iota_1}(\bar u)a(v)=
b^{\iota_1}(v)b(\bar u)\tilde d(u,v)+c^{\iota_1}(v)d(\bar u)\tilde a(u,v),
$$
$$ 
a(u,v)d^{\iota_1}(\bar u)c(v)+d(u,v)b^{\iota_1}(\bar u)b(v)=
a^{\iota_1}(v)a(\bar u)\tilde d(u,v)+d^{\iota_1}(v)c(\bar u)\tilde a(u,v),
$$
which represents the non-trivial part of $XYZ$ model. In
the $XXZ$ case, when $d(u)=0$, this set of equations disappears.

Our aim is now to find a solution of equations (\ref{i}), (\ref{i2}), 
(\ref{i3}) with non-trivial $\iota_1$ operation. 
It can be seen easily that by defining
$$
a^{\iota_1}(u)=a(u),\qquad \tilde a(u,v)=a(u,v),
$$
\begin{equation}
\label{1t}
b^{\iota_1}(u)=-b(u),\qquad \tilde b(u,v)=-b(u,v)
\end{equation}
$$
c^{\iota_1}(u)=c(u),\qquad \tilde c(u,v)=c(u,v)
$$
$$
d^{\iota_1}(u)=-d(u),\qquad \tilde d(u,v)=-d(u,v)
$$
the eqs. (\ref{i}) obeyed identically, while the eqs. 
(\ref{i2}) and (\ref{i3}) are reducing to the corresponding equations of 
the staggered case. Therefore the ordinary choice of parametrisation
of $a(u), b(u), c(u)$ and $d(u)$ via the Jacobi elliptic
functions 

$$
a(u)=\sn(u+\eta)=a^{\iota_1}(u),\qquad\qquad a(u,v)=
\sn(\bar u-v+\eta)=\tilde a(u,v),
$$
$$
b(u)=\sn u=-b^{\iota_1}(u),\qquad\qquad\quad b(u,v)=
\sn(\bar u-v)=-\tilde b(u,v),
$$
\begin{equation}
\label{t1}
c(u)=\sn\eta=c^{\iota_1}(u),\qquad\quad\qquad\qquad c(u,v)=
\sn\eta=\tilde c(u,v),
\end{equation}
$$
d(u)=k\sn\eta\sn u\sn(u+\eta)=-d^{\iota_1}(u),
$$
$$
d(u,v)=k\sn\eta \sn(\bar u-v)\sn(\bar u-v+\eta)=-\tilde d(u,v),
$$
fulfils the remaining equations (\ref{i2}) and (\ref{i3}).

Let us analyse now the second set of $YBE$'s (\ref{eq:YBE2}). 
First we conclude immediately from the solution (\ref{1t})
of the previous set of $YBE$'s that the operation  $\tilde{}~$  
is simply coinciding with the operation $\iota_1$.
Then it is easy to see, that if we define $\iota_2$
operation in (\ref{eq:YBE2}) as

\begin{equation}
\label{jj}
R^{\iota_2}(u)=R^{\iota_1}(-u) \;,
\end{equation}
then the second set (\ref{eq:YBE2}) of $YBE$'s 
coincides with the first set (\ref{eq:YBE1}) after additional
action on it by $\iota_1$. This means that the relation (\ref{jj})
is ensuring the fulfilment of the second set (\ref{eq:YBE2})
of $YBE$'s. Therefore we have

\begin{equation}
\label{jjj}
a^{\iota_2}(u)=a(-u),\quad b^{\iota_2}(u)=-b(-u),\quad c^{\iota_2}(u)=c(-u),
\quad d^{\iota_2}(u)=-d(-u).
\end{equation}

It is now necessary to emphasise that, as in ordinary case, the
parameters $\Delta$ and $k$ defined as
\begin{equation}
\label{delta-k}
\Delta=\frac{a^2+b^2-c^2-d^2}{2ab},\qquad \qquad k \sn^2\eta=\frac{cd}{ab}
\end{equation}
are constants (as well as $\Delta^{\iota_1}$ and $k^{\iota_1}$). They are
defining the anisotropy of the model 
in the $z$ and $y$ directions. As one can see, now, due to solution 
(\ref{t1}), the anisotropy parameter $\Delta$ is staggered along the
chain and time directions.

Hence, we have found the solution of staggered $YBE$'s 
(\ref{eq:YBE1}), (\ref{eq:YBE2}) and can calculate
now the Hamiltonian of corresponding integrable model.

\section{The Transfer matrix and the Hamiltonian }
\setcounter{equation}{0}

Having the solution of graded $YBE$'s one can start the
calculation of the monodromy matrix and the Hamiltonian.
According to formula (\ref{TT}) the monodromy matrix of the
model is
\begin{eqnarray}
\label{mm}
&&\hspace{-20mm}
T^{ab,\;\; i_1...i_{2N}}_{cd,\;\; k_1...k_{2N}}(u,\theta)\equiv
T^{\;\;\,\;a ,\;\; i_1...i_{2N}}_{0\;\;c,\;\; j_1...j_{2N}}(u,
\theta) \;
T^{\;\;\,\;b ,\;\; j_1...j_{2N}}_{1\;\;d,\;\; k_1...k_{2N}
}(u,\theta)\nonumber\\[2mm]
&=&\left(R^{\iota_1\;\;\, a\;\, i_1}_{0,1\;\;a_1j_1}(\bar u)
\;R^{\iota_1\iota_2\;\; a_1i_2}_{0,2\;\;\;\;\;a_2j_2}(u)\;R^{\iota_1\;\;\, 
a_2i_3}_{0,3\;\;a_3j_3}(\bar u)...\;R^{\iota_1\iota_2\;\; a_{2N-1}i_{2N}}_
{0,2N\;\;\;c\;\;\;\;\;j_{2N}}(u)\right)\nonumber\\[2mm]
&&\left(R^{\;\;\;\;\;\;\, b\; j_1}_{0^\prime 1\;\;b_1k_1}(u)
\;R^{\iota_2\;\;\;\, b_1j_2}_{0^\prime ,2\;\;b_2k_2}(\bar u) 
\;R^{\;\;\;\;\;\;\;\, b_2j_3}_{0^\prime ,3\;\;b_3k_3}(u)...
\;R^{\iota_2\;\;\; b_{2N-1}j_{2N}}_{0^\prime ,2N\;\;d\;\;\;k_{2N}}
(\bar u)\right)\\[2mm]
&=&\left(R^{\iota_1\;\;\; a\;\, i_1}_{0,1\;\;a_1j_1}(\bar u)
\;R^{\;\;\;\;\;\;\;\, b\;\, j_1}_{0^\prime ,1\;\;b_1k_1}(u) 
\;R^{\iota_1\iota_2\;\; a_1i_2}_{0,2\;\;\;\;\;a_2j_2}(u)
\;R^{\iota_2\;\;\;\, b_1j_2}_{0^\prime ,2\;\;b_2k_2}(\bar u)\right)...\nonumber
\\[2mm]
&&
\left(R^{\iota_1\;\;\;\;\;\; \;\;\;\;
a_{2N-2}i_{2N-1}}_{0,2N-1\;\;a_{2N-1}j_{2N-1}}(\bar u)
\;R^{\;\;\;\;\;\;\;\;\;\;\;\;\; b_{2N-2}j_{2N-1}}_{0^\prime
,2N-1\;\;b_{2N-1}k_{2N-1}}(u)\;R^{\iota_1\iota_2\;\;\, a_{2N-1}i_{2N}}_
{0,2N\;\;\;c\;\;\;\;\;\;\; j_{2N}}(u)   \;R^{\iota_2\;\;\; b_{2N-1}
j_{2N}}_{0^\prime ,2N\;d\;\;\;\;k_{2N}}(\bar u)\right).
\end{eqnarray}

Lets write now an explicit formula for the $R$-matrix
in terms of Pauli matrices
\begin{eqnarray}
\label{rmat}
R^{\;\;\;\; ai}_{0,r\;cj}(u)=
a(u)\left(\sigma^{\;\;\; a}_{0\; c}
\sigma^{\;\;\; i}_{r\; j}+
\bar\sigma^{\;\;\; a}_{0\; c}\bar\sigma^{\;\;\; i}_{r\; j}\right)
+b(u)\left(\sigma^{\;\;\; a}_{0\; c}\bar\sigma^{\;\;\; i}_{r\; j}+
\bar\sigma^{\;\;\; a}_{0\; c}\sigma^{\;\;\; i}_{r\; j}\right)+\nonumber\\
c(u)\left(\sigma^{+\;\; a}_{0\;\;\, c}\sigma^{-\;\; i}_{r\;\;\, j}+
\sigma^{-\;\; a}_{0\;\;\, c}\sigma^{+\;\; i}_{r\;\;\, j}\right)+
d(u)\left(\sigma^{+\;\; a}_{0\;\;\, c}\sigma^{+\;\; i}_{r\;\;\, j}+
\sigma^{-\;\; a}_{0\;\;\, c}\sigma^{-\;\; i}_{r\;\;\, j}\right),
\end{eqnarray}
where $0$ and $r$ refer to auxiliary and quantum spaces 
respectively and
$$
\sigma\equiv\frac 12\left(\sigma^4+\sigma^z\right), \quad
\bar\sigma\equiv\frac 12\left(\sigma^4-\sigma^z\right), \quad
\sigma^+\equiv\frac 12\left(\sigma^x+i\sigma^y\right), \quad
\sigma^-\equiv\frac 12\left(\sigma^x-i\sigma^y\right).
$$

Then transfer matrix is defined as trace of monodromy matrix 
(\ref{mm}) over auxiliary spaces and at the zero value of
the spectral parameter $u$ is equal to
\begin{eqnarray}
\tau^{i_1...i_{2n}}_{k_1...k_{2n}}(0,\theta) &\equiv&
T^{ab,i_1...i_{2N}}_{ab,k_1...k_{2N}}(0,\theta) \\
\label{tm}
&=&
tr_0tr_{0^\prime}\prod^N_{r=1}\left(R^{\iota_1\; a_{2r-1}i_{2r-1}}_
{\;\;\;\;\; a_{2r}\;\;\; j_{2r-1}}(\theta)R^{b_{2r-1}j_{2r-1}}_
{b_{2r}\;\;\; k_{2r}}(0)R^{a_{2r}\;\;\; i_{2r}}_{a_{2r+1}j_{2r}}
(0)R^{\iota_1\; b_{2r}\;\, j_{2r}}_{\;\;\; b_{2r+1}k_{2r}}(-\theta)
\right)
\\
&=&
tr_0tr_{0^\prime}\prod^N_{r=1}\left(\sn^2\eta(\sn^2\eta-\sn^2
\theta)P_{0,2r}P_{0^\prime,2r-1}\right)
\\
&=&
\left(\sn^2\eta(\sn^2\eta-\sn^2\theta)\right)^N
\delta^{i_1}_{k_3}\delta^{i_2}_{k_4}...\delta^{i_{2N-2}}_{k_{2N}}
\delta^{i_{2n-1}}_{k_1}\delta^{i_{2N}}_{k_2},
\end{eqnarray}
i.e. up to overall multiplier it is becoming an operator that shifts
along the chain by two units, i.e. a translation operator. Here we
have used 
$$
R^{\;\;\;\; ai}_{0,r\;cj}(0)=\sn\eta\cdot\delta^i_c\delta^a_j
\equiv \sn\eta\cdot P_{0,r}
$$
where $P_{0,r}$ is an operator permuting auxiliary ($0$-th) and quantum 
($r$-th) spaces. We also used the relation (unitarity property of $R$)
$$
R^{\iota_1\;a_{2r-1}i_{2r-1}}_{\;\;\;\; a_{2r}\;\;\;b_{2r}}(\theta)
R^{\iota_1\;b_{2r}a_{2r}}_{\;\; b_{2r+1}k_{2r}}(-\theta)=
\left(\sn^2\eta-\sn^2\theta\right)\delta^{a_{2r-1}}_{k_{2r}}
\delta^{i_{2r-1}}_{b_{2r+1}},
$$
which holds due to following identities for elliptic functions:
$$
a(u)a(-u)+d(u)d(-u)=\sn^2\eta-\sn^2u=b(u)b(-u)+c(u)c(-u)
$$
$$
a(u)d(-u)+d(u)a(-u)=0=b(u)c(-u)+c(u)b(-u).
$$

The functions $a(u)$, $b(u)$, $c(u)$ and $d(u)$ are given by (\ref{t1})
Now we can turn to calculation of the Hamiltonian
\begin{equation}
\label{ham}
\left.\frac{d}{du}\log\tau \right|_{u=0}=\sn^2\eta(\sn^2\eta-\sn^2u) \;\;
\cH^{j_1...j_{2N}}_{k_1...k_{2N}} 
\; P^{i_3...i_{2N}\; i_1\;i_2}_{j_1...j_{2N-1}j_{2N}} 
\end{equation}

One can see, that the Hamiltonian of the model is reduces to the sum of
contributions from the derivatives of quartic products 
of neighbour $R$-matrices inside of 
brackets in (\ref{tm}) after the  permutation of indices. 
It appears that the differentiation of two 
terms in the bracket again contributes to the identity operator and
provides a constant contribution to Hamiltonian. 
Therefore the non-trivial contribution looks like
\begin{equation}
\label{ham1}
\left(R^{\iota_1\;i_2i_3}_{\;\;\;\;\; a_2a_3}(\theta)R^{\prime\;i_1
a_3}_{\;\;a_1k_1}(0)R^{\iota_1\;a_1a_2}_{\;\;\;\;\; k_3k_2}(-\theta)
\delta^{i_4}_{k_4}...\delta^{i_{2N}}_{k_{2N}}-R^{\iota_1\;i_2i_3}_
{\;\;\;\;\; a_2a_3}(\theta)R^{\prime\;a_2i_4}_{\;\;k_4a_4}(0)R^
{\iota_1\;a_3a_4}_{\;\;\;\;\; k_3k_2}(-\theta)\delta^{i_5}_{k_5}...
\delta^{i_1}_{k_1}\right)+...\nonumber\\
\end{equation}

After some calculations by use of properties of elliptic functions
and (\ref{delta-k}) one can find the following
expression for the Hamiltonian (without the constant part)
\begin{eqnarray}
\label{hami}
\cH&=&\sum^N_{r=1}\cH_{2r-1,2r,2r+1,2r+2},
\\
&&\hspace{-13mm}
\sn^2\eta(\sn^2\eta-\sn^2u)
\cH_{2r-1,2r,2r+1,2r+2}=
\\&&\hspace{-10mm}
\frac{1}{2}\left(a(u)a(-u)-c(u)c(-u)\right)a^{\prime}(0)
\left[\sigma^z_{2r-1}\sigma^z_{2r+1}-\sigma_{2r}^z\sigma^z_{2r+2}\right]
\nonumber\\&&\hspace{-10mm}
+\frac{b(u)}{2}\left(a(u)-a(-u)\right)
\left(b^{\prime}(0)-k c^2(u)d^{\prime}(0)\right)
\left[\sigma^+_{2r-1}\sigma^-_{2r+1}+\sigma^-_{2r-1}\sigma^+_
{2r+1}-\sigma^+_{2r}\sigma^-_{2r+2}-\sigma^-_{2r}\sigma^+_{2r+2}\right]
\nonumber\\&&\hspace{-10mm}
+\frac{b(u)}{2}\left(a(u)-a(-u)\right)
\left(d^{\prime}(0)-k c^2(u)b^{\prime}(0)\right)
\left[\sigma^+_{2r-1}\sigma^+_{2r+1}+\sigma^-_{2r-1}\sigma^-_
{2r+1}-\sigma^+_{2r}\sigma^+_{2r+2}-\sigma^-_{2r}\sigma^-_{2r+2}\right]
\nonumber\\&&\hspace{-10mm}
-b(u)c(u)a^{\prime}(0)
\left[\sigma^z_{2r}(\sigma^+_{2r+1}\sigma^-_{2r+2}-
\sigma^-_{2r+1}\sigma^+_{2r+2})+
(\sigma^+_{2r-1}\sigma^-_{2r}-\sigma^-_{2r-1}\sigma^+_{2r})\sigma^z_{2r+1}
\right]
\nonumber\\&&\hspace{-10mm}
-\frac{b(u)}{2}\left(a(u)+a(-u)\right)
\left(b^{\prime}(0)-k c^2(u)d^{\prime}(0)\right)
\left[\sigma^z_{2r+1}(\sigma^+_{2r}\sigma^-_{2r+2}-
\sigma^-_{2r}\sigma^+_{2r+2})+
(\sigma^+_{2r-1}\sigma^-_{2r+1}-\sigma^-_{2r-1}\sigma^+_{2r+1})\sigma^z_{2r}
\right]
\nonumber\\&&\hspace{-10mm}
+\frac{c(u)}{2}\left(a(u)-a(-u)\right)
\left(b^{\prime}(0)-k b^2(u)d^{\prime}(0)\right)
\left[\sigma^z_{2r-1}(\sigma^+_{2r}\sigma^-_{2r+1}-
\sigma^-_{2r}\sigma^+_{2r+1})+
(\sigma^+_{2r}\sigma^-_{2r+1}-\sigma^-_{2r}\sigma^+_{2r+1})\sigma^z_{2r+2}
\right]
\nonumber\\&&\hspace{-10mm}
+a(u)d(u)a^{\prime}(0)
\left[\sigma^z_{2r}(\sigma^+_{2r+1}\sigma^+_{2r+2}-
\sigma^-_{2r+1}\sigma^-_{2r+2})-
(\sigma^+_{2r-1}\sigma^+_{2r}-\sigma^-_{2r-1}\sigma^-_{2r})\sigma^z_{2r+1}
\right]
\nonumber\\&&\hspace{-10mm}
+\frac{b(u)}{2}\left(a(u)+a(-u)\right)
\left(d^{\prime}(0)-k c^2(u)b^{\prime}(0)\right)
\left[\sigma^z_{2r+1}(\sigma^+_{2r}\sigma^+_{2r+2}-
\sigma^-_{2r}\sigma^-_{2r+2})-
(\sigma^+_{2r-1}\sigma^+_{2r+1}-\sigma^-_{2r-1}\sigma^-_{2r+1})\sigma^z_{2r}
\right]
\nonumber\\&&\hspace{-10mm}
+\frac{c(u)}{2}\left(a(u)-a(-u)\right)
\left(d^{\prime}(0)-k b^2(u)b^{\prime}(0)\right)
\left[\sigma^z_{2r-1}(\sigma^+_{2r}\sigma^+_{2r+1}-\sigma^-_{2r}
  \sigma^-_{2r+1})-  
(\sigma^+_{2r}\sigma^+_{2r+1}-\sigma^-_{2r}\sigma^-_{2r+1})\sigma^z_{2r+2}
\right]
\nonumber
\end{eqnarray}
This expression for the Hamiltonian due to next to nearest
neighbour interactions can be easier understood as
one, written on a zig-zag chain.

%%%%%%%%%%%%%%%%%%%%%%%%%%%%%%%%%%%%%%%%%%%%%%%%%%%%%%%%%%%%%%%%
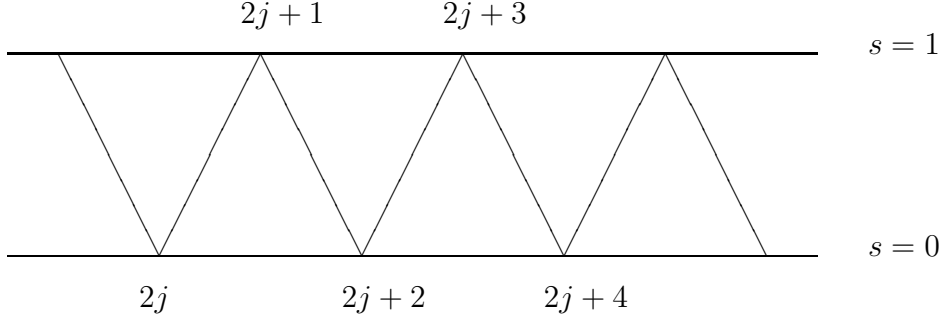
\begin{figure}[h]
  \begin{center}
    \leavevmode
\setlength{\unitlength}{0.0015cm}
\begingroup\makeatletter\ifx\SetFigFont\undefined
% extract first six characters in \fmtname
\def\x#1#2#3#4#5#6#7\relax{\def\x{#1#2#3#4#5#6}}%
\expandafter\x\fmtname xxxxxx\relax \def\y{splain}%
\ifx\x\y   % LaTeX or SliTeX?
\gdef\SetFigFont#1#2#3{%
  \ifnum #1<17\tiny\else \ifnum #1<20\small\else
  \ifnum #1<24\normalsize\else \ifnum #1<29\large\else
  \ifnum #1<34\Large\else \ifnum #1<41\LARGE\else
     \huge\fi\fi\fi\fi\fi\fi
  \csname #3\endcsname}%
\else
\gdef\SetFigFont#1#2#3{\begingroup
  \count@#1\relax \ifnum 25<\count@\count@25\fi
  \def\x{\endgroup\@setsize\SetFigFont{#2pt}}%
  \expandafter\x
    \csname \romannumeral\the\count@ pt\expandafter\endcsname
    \csname @\romannumeral\the\count@ pt\endcsname
  \csname #3\endcsname}%
\fi
\fi\endgroup
{\renewcommand{\dashlinestretch}{30}
%%%%%%%%%%%%%%%%%%%%%%%%%%%%%%%%%%%%%%%%%%%%%%%%%%%%%%%%%%%%%%%%
\begin{picture}(7954,2733)(0,-10)
\drawline(462,2304)(1362,504)(2262,2304)
        (3162,504)(4062,2304)(4962,504)
        (5862,2304)(6762,504)
\drawline(12,2304)(7212,2304)
\drawline(12,504)(7212,504)
\put(1182,54){\makebox(0,0)[lb]{\smash{{{\SetFigFont{12}{14.4}
{\rmdefault}$2j$}}}}}
\put(2982,54){\makebox(0,0)[lb]{\smash{{{\SetFigFont{12}{14.4}
{\rmdefault}$2j+2$}}}}}
\put(4782,54){\makebox(0,0)[lb]{\smash{{{\SetFigFont{12}{14.4}
{\rmdefault}$2j+4$}}}}}
\put(2082,2574){\makebox(0,0)[lb]{\smash{{{\SetFigFont{12}{14.4}
{\rmdefault}$2j+1$}}}}}
\put(3882,2574){\makebox(0,0)[lb]{\smash{{{\SetFigFont{12}{14.4}
{\rmdefault}$2j+3$}}}}}
\put(7662,2304){\makebox(0,0)[lb]{\smash{{{\SetFigFont{12}{14.4}
{\rmdefault}$s=1$}}}}}
\put(7662,504){\makebox(0,0)[lb]{\smash{{{\SetFigFont{12}{14.4}
{\rmdefault}$s=0$}}}}}
\end{picture}
} 
\end{center}
\caption{Zig-zag ladder chain}
\label{fig}
\end{figure}
%%%%%%%%%%%%%%%%%%%%%%%%%%%%%%%%%%%%%%%%%%%%%%%%%%%%%%%%%%%%%%%%
Let us introduce two chains
consisting of the even and odd sites of the original chain and
label them by $s = 0$ and $ 1$ correspondingly. 
Make now zig-zag rungs as it is shown in the Fig.1 and introduce
the following labelling of Pauli matrices 
\begin{equation}
\label{CC}
\vec\sigma_{j,s}=\vec\sigma_{2j+s}, \qquad \qquad s = 0,1
\end{equation}
Then, substituting the expression (\ref{t1}) for $a(u),b(u),c(u),d(u)$
to (\ref{hami}) and after some interesting cancellations of some terms
due to identities on elliptic functions, 
one can write the following zig-zag ladder Hamiltonian
\begin{eqnarray}
\label{hh}
&&\hspace{-20mm}
(1-k^2 \sn^2\eta \sn^2\theta)(\sn^2\eta-\sn^2\theta) \cH_{j,s}=\nonumber\\
&=&\frac{(-1)^{s+1}}{2}\sn^2\theta \cn\eta \dn\eta (1-k^2 \sn^4\eta)
\left[\sigma^1_{j,s}\sigma^1_{j+1,s} + 
\sigma^2_{j,s}\sigma^2_{j+1,s}-\sigma^3_{j,s}\sigma^3_{j+1,s}  
\right]\nonumber\\
&+&\hat\epsilon_s^{abc} \sigma_{j,s}^a \sigma_{j,s+1}^b \sigma_{j+1,s}^c
+\hat\tau_s^{abc} \sigma_{j,s}^a \sigma_{j,s+1}^b
\sigma_{j+1,s}^c,\qquad\;\; s=0,1 
\end{eqnarray}
where the anisotropic antisymmetric tensors $\hat\epsilon_s^{abc}$
and $\hat\tau_s^{abc}$ are defined as follows:
\begin{eqnarray}
\label{ee1}
&&\hat\epsilon_0^{3+-} = -\hat\epsilon_0^{3-+}=-
\hat\epsilon_0^{+-3} = \hat\epsilon_0^{-+3}
=-\hat\epsilon_1^{3+-} = \hat\epsilon_1^{3-+}=
\hat\epsilon_1^{+-3} = -\hat\epsilon_1^{-+3} 
%%= \nonumber\\&& \hspace{4cm}
=-\sn\theta \sn\eta \cn\eta \dn\eta(1-k^2 \sn^2\eta \sn^2\theta),\nonumber\\
&&\hat\epsilon_0^{+3-} = -\hat\epsilon_0^{-3+}=
\hat\epsilon_1^{+3-} = -\hat\epsilon_1^{-3+}
=\sn\eta \sn\theta \cn\theta \dn\theta (1-\sn^4\eta) 
\end{eqnarray}
and
\begin{eqnarray}
\label{ee2}
&&\hat\tau_0^{3++} = -\hat\tau_0^{3--}=
\hat\tau_0^{++3} = -\hat\tau_0^{--3}
%%=\\&&
=-\hat\tau_1^{3++} = \hat\tau_1^{3--}=
-\hat\tau_1^{++3} = \hat\tau_1^{--3}=
-k \sn\eta \sn\theta \cn\eta \dn\eta(\sn^2\eta- \sn^2\theta),\nonumber\\
&&\hat\tau_0^{+3+} = \hat\tau_0^{-3-}=
\hat\tau_1^{+3+} = \hat\tau_1^{-3-}=0 \;.
\end{eqnarray}
This tensors define the interaction terms of topological character.
The $XXZ$ limit ($k$=0) of the Hamiltonian (\ref{hh}) coincides with
the corresponding expression found in \cite{APSS}.

\section*{Acknowledgement}
\indent

The authors D.K., M.M., and A.S. acknowledge INTAS grants 00-390 and
00-561 for the partial financial support.

\end{document}